# Low-energy collisions between carbon atoms and oxygen molecules in a magnetic trap

Michael Karpov*, Martin Pitzer*, Yair Segev, Julia Narevicius and Edvardas Narevicius

Department of Chemical and Biological Physics, Weizmann Institute of Science, Rehovot 7610001, Israel.

**Abstract**

Trapping of atoms and molecules in electrostatic, magnetic and optical traps has enabled studying atomic and molecular interactions on a timescale of many seconds, allowing observations of ultra-cold collisions and reactions. Here we report the first magnetic deceleration and trapping of neutral carbon atoms in a static magnetic trap. When co-trapping the carbon atoms with oxygen molecules in a superconducting trap, the carbon signal decays in a non-exponential manner, consistent with losses resulting from atom-molecule collisions. Our findings pave the way to studying both elastic and inelastic collisions of species that cannot be laser cooled, and specifically may facilitate the observation of reactions at low temperatures, such as C + $O_2 \rightarrow$ CO + O, which is important in interstellar chemistry.

## 1. Introduction

Collisions between atoms and molecules play a central role in many physical processes, including chemical reactions and scattering phenomena. Experimental tests at low collision energies where only a few quantum states contribute are useful for microscopic understanding of collision processes and for benchmarking state-of-the-art theoretical models [1]. Such studies are also crucial for improving collisional cooling techniques such as sympathetic cooling, where laser-cooled atoms lower the temperature of other species through energy transfer via collisions [2,3], or buffer-gas cooling, where the species of interest thermalizes with a cold gas, usually helium or neon [4].

Many experiments have reported various types of collisions in mixtures containing cold molecules. Some of the lowest collision temperatures to date have been achieved in experiments that rely on laser cooling as an initial stage, which is used to directly cool atoms and molecules [5,6]. These laser-cooled atoms can be assembled into molecules via several techniques, such as magneto- and photo-association. Bimolecular collisions have been reported in a number of experiments [7-10], and collisions between atoms and molecule have been observed with molecules assembled from alkali atoms [6,11,12,13].

Many atoms and molecules are not amenable to laser cooling, in particular those of interest to naturally occurring chemical reactions such as carbon, nitrogen or oxygen. Most studies with such species have been performed at collision temperatures above 10 K using crossed molecular beam setups [14-16]. During the last decade, merged beams have allowed lowering the relative kinetic energies between collision partners to values corresponding to the subkelvin regime [17,18]. However, investigation of small cross-section processes remains challenging in these setups due to the short interaction times and the low densities of collision partners. In order to extend the experimental timescales to the order of seconds, the colliding particles can be confined using electrostatic [19,20], magnetic [21,22] or optical [9] traps. Only a few experiments have demonstrated collisions between trapped partners that cannot be laser cooled. For example, buffer-gas cooling enabled the observation of collisions between nitrogen atoms and NH molecules, three-body collisions between Ag atoms and $^3$He [23,24], and reactive collisions between Li atoms and CaH [25]. Stark deceleration and subsequent electrostatic trapping of $ND_3$ [26] was used to observe inelastic collisions with magnetically trapped rubidium atoms [27]. In most atom-molecule collision experiments the majority species was atoms. Recently, we trapped oxygen molecules using magnetic fields, measuring bimolecular collisions as well as collisions between oxygen molecules and lithium atoms [28].

Here we extend our method and simultaneously trap two species that are not amenable to laser cooling, measuring collisions between neutral carbon atoms and molecular oxygen. Cold oxygen molecules have been trapped in several recent experiments [29,30]. We demonstrate the first trapping of carbon atoms, alongside oxygen molecules. This approach may allow us to study the barrierless reaction of atomic carbon with molecular oxygen $C(^3P) + O_2(^3\Sigma^-_g) \rightarrow CO(^1\Sigma^+) + O(^1D)$ at subkelvin temperatures. This reaction has received much attention in the context of chemistry of the interstellar medium, and has been studied theoretically [31] and experimentally [32] at temperatures above 15 K.

## 2. Methods

The starting point of the experiment, depicted in Figure 1, is the production of an ensemble of cold oxygen molecules seeded in a krypton carrier gas, emitted as a supersonic beam from a pulsed Even-Lavie valve [33]. Supersonic expansion cools the oxygen molecules, in the external as well as the internal degrees of freedom, through collisions [34]. Carbon atoms are entrained into the molecular beam by sweeping the beam through a sparse plume of laser-ablated carbon near the valve. This plume is generated by focusing nanosecond pulses of 30 mJ at 532 nm onto a solid graphite rod. The mixed beam has a temperature on the order of 1K, and its average velocity of 375 m/s is obtained by cooling the valve to 165 K. In order to magnetically trap the particles with practical static fields, we must first decelerate them to velocities of about 10 m/s. Our moving-trap Zeeman decelerator [35] is capable of adiabatically slowing ensembles of paramagnetic species using time-varying magnetic fields, including co-trapping of mixtures in a superconducting magnetic trap 0.8T deep [28], which corresponds to 0.5K deep for carbon and 0.8K for oxygen. We detect carbon atoms in the $^3P_1$ state by using 2+1 Resonance Enhanced Multi-Photon Ionization (REMPI) with ~10 mJ nanosecond pulses of several nanoseconds duration at 280.3142 nm [36] and extracting the resulting ions to a micro-channel plate (MCP) detector using an array of electrodes. Molecular oxygen is similarly detected using a 2+1 REMPI scheme, with pulses at around 287 nm [37].

The focused laser beam enters perpendicularly to the deceleration axis and ionizes a cylindrical volume along the center of the trap, as its Rayleigh range is comparable to the trap's width. Therefore, the measurement is a column integration of the density along the axis of the laser beam [29]. The measurements are performed for different delays between loading and ionization in order to estimate the fraction of particles remaining in the trap as a function of the holding time. A decay in the amount of trapped carbon can be attributed to collisions with background gas and with other trapped particles, specifically the denser oxygen molecules. To differentiate between these contributions, we compare the decay of carbon atoms loaded alongside oxygen molecules from a mixed krypton/oxygen beam to the decay of carbon atoms trapped without molecules by entrainment in a pure krypton beam.

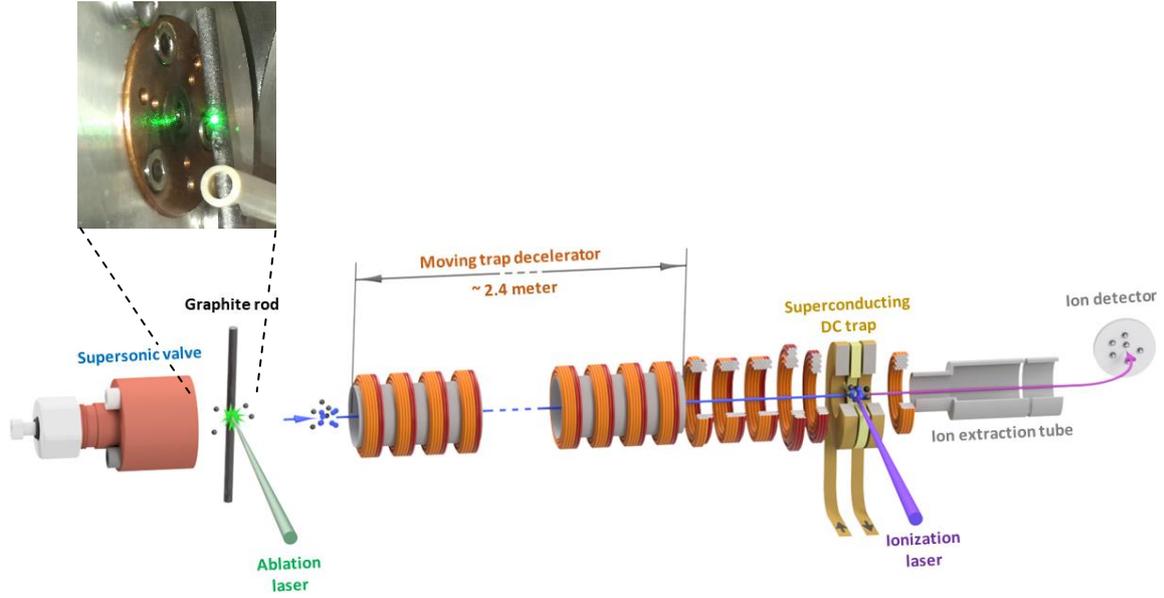

**Figure 1**. Schematic diagram of the experimental setup. A beam of mixed oxygen and krypton gas is emitted from a pulsed valve. Carbon atoms are entrained in the beam by laser-ablating a solid graphite target near the nozzle. After the beam enters the decelerator, the oxygen and carbon are magnetically slowed down to a trappable velocity and loaded into a superconducting trap. Arrows on the DC trap indicate the current direction. After the desired trapping time, the atoms or molecules are detected by ionization using a 2+1 REMPI process and extraction towards an MCP detector using an array of electrodes.

3. **Results**

The decay of the column-integrated carbon signal as a function of the holding time is plotted in Figure 2. The measured signal of carbon atoms decelerated from a pure krypton beam is well described by fitting to the model $dn_C/dt = -\alpha n_C$, where $n_C$ is the carbon column-integrated density signal. This model corresponds to a single exponential decay due to collisions with background gas, with a lifetime of $\tau = \alpha^{-1}$ = 18 seconds.

Introducing oxygen molecules into the supersonic beam and co-trapping them with the carbon atoms results in a clear deviation of the measured carbon signal from an exponential decay, indicating that in this case the dynamics is not governed by collisions with background gas anymore. Instead, the dynamics can be described by the model $dn_C/dt = -\beta n_{O_2} n_C$, which takes into account the fact that the rate of decay of carbon atoms depends on the density of the oxygen molecules. The presence of oxygen molecules speeds the decay of the carbon signal at early times by a factor of five, resulting in a two-body lifetime [11] $\tau_{2B,C-O_2} = \left(\beta \cdot n_{O_2}(0)\right)^{-1}$ = 3.5s. Although both decays are normalized to an initial value of one, note that the signal of trapped carbon at the loading time is 25% lower when co-trapped alongside oxygen.

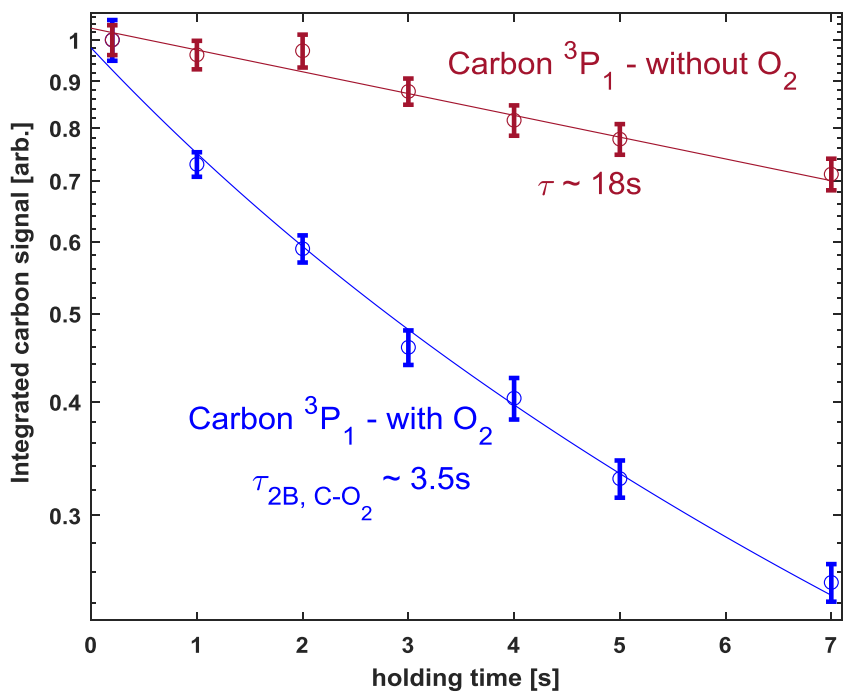

**Figure 2.** Measurement of the column-integrated density of trapped carbon $^3P_1$ atoms, held with (red) and without (blue) oxygen molecules, as a function of the holding time in the static trap, where each point was obtained by averaging over 50 repetitions. When oxygen molecules are not present in the trap, the carbon signal decays exponentially with a 1/e lifetime of 18 seconds. Co-trapping the atoms with oxygen causes the carbon signal to decay non-exponentially with a much shorter two-body lifetime of 3.5 seconds. Although both decays are normalized to an initial value of one, the signal of trapped carbon at the loading time is 25% lower when co-trapped alongside oxygen. Error bars represent on standard error.

## Discussion

The nature of this non-exponential decay can be attributed to three possible collision mechanisms: elastic, inelastic and reactive collisions. Elastic collisions do not change the quantum state of the atoms, therefore usually preserving their number in the trap and redistributing the energy between the colliding partners. However, an elastic collision between two particles near the edge of the trap can lead to ejection (evaporative loss) due to momentum transfer pushing at least one particle into an untrapped trajectory. Carbon is more susceptible to evaporative losses following elastic collisions with $O_2$ due to its lighter mass that leads to higher recoil velocities. Inelastic collisions may change the quantum state of the particles into an untrapped state, while a probable reactive loss mechanism is the C+ $O_2 \rightarrow$ CO + O reaction. Differentiating between these processes requires further investigation. The difficulty is that the reaction is exothermic [32], creating high-energy products that cannot be confined by our trap. Thus, a more sensitive detection scheme needs to be implemented to probe the products that continuously leave the trap. Detection of atomic oxygen is challenging, and requires single photon VUV ionization from the full volume of the trap [38]. An additional approach may be to employ other stable oxygen isotopologues such as $^{16}O^{18}O$ or $^{17}O^{17}O$, which may have different collision properties [39] that can help differentiate between elastic and inelastic collisions.

This work paves the way for studying collisions of paramagnetic species that are not amenable to laser cooling with the prospects of identifying the role of elastic and inelastic contributions to the dynamics. Specifically, directly observing products of the C+ $O_2 \rightarrow$ CO + O reaction, which is important in modelling chemistry in interstellar clouds, would enable studying this reaction at sub-Kelvin temperatures.

**Acknowledgments**
We thank A. Kuprienko of the Weizmann Chemical Research Support and H. Sade of the Weizmann Institute CNC Section for assistance in designing and manufacturing the experiment components. We acknowledge funding from the European Research Council and the Israel Science Foundation.